\documentclass[pra,floatfix,preprint,11pt]{revtex4}

\def\Leff{L_{\mathrm{eff}}}
\def\bEp{\mathbf{E}_p}
\def\bEe{\mathbf{E}_e}
\def\bAp{\mathbf{A}_p}
\def\bAe{\mathbf{A}_e}
\def\bk{\mathbf{k}}
\def\br{\mathbf{r}}
\def\ceff{\chi_{\mathrm{eff}}}
\def\OV{\Omega_{v}}
\def\Da{\Delta \alpha}

\def\bP{\mathbf{P}}\def\bE{\mathbf{E}}

\usepackage{amsmath}
\usepackage{graphicx}
\usepackage{bm}
\usepackage{hyperref}

\begin{document}

\title{Absolute measurement of the ultrafast nonlinear electronic and rovibrational response in H$_2$ and D$_2$}

\author{J. K. Wahlstrand}
\affiliation{Institute for Research in Electronics and Applied Physics, University of Maryland, College Park, MD 20742}
\author{S. Zahedpour}
\affiliation{Institute for Research in Electronics and Applied Physics, University of Maryland, College Park, MD 20742}
\author{Y.-H. Cheng}
\altaffiliation[Present address: ]{National Institute for Standards and Technology, Gaithersburg, MD}
\affiliation{Institute for Research in Electronics and Applied Physics, University of Maryland, College Park, MD 20742}
\author{J. P. Palastro}
\altaffiliation[Present address: ]{Naval Research Laboratory, Washington, DC}
\affiliation{Institute for Research in Electronics and Applied Physics, University of Maryland, College Park, MD 20742}
\author{H. M. Milchberg}
\email{milch@umd.edu}
\affiliation{Institute for Research in Electronics and Applied Physics, University of Maryland, College Park, MD 20742}



\begin{abstract}
The electronic, rotational, and vibrational components of the ultrafast optical nonlinearity in H$_2$ and D$_2$ are measured directly and absolutely at intensities up to the ionization threshold of $\sim$10$^{14}$ W/cm$^2$. As the most basic nonlinear interactions of the simplest molecules exposed to high fields, these results constitute a benchmark for high field laser-matter theory and simulation. 
\end{abstract}

\maketitle

\section{Introduction}

The high field nonlinear polarizability of diatomic molecules is a topic of practical and fundamental importance. From optical frequency conversion \cite{kim_coherent_2008,popmintchev_bright_2012} to simulations and experiments in high intensity propagation \cite{couairon_femtosecond_2007,berge_ultrashort_2007,kolesik_modeling_2014,penano_propagation_2004,varma_trapping_2008,chen_direct_2010,palastro_compression_2012}, understanding the atomic and molecular behaviour under high laser fields is crucial for applications. Unlike in monatomic gases, the nonlinear molecular response cannot be characterized by a single coefficient at a given optical frequency: processes at several timescales contribute, and so the response depends on pulse duration \cite{varma_trapping_2008,chen_direct_2010,nibbering_determination_1997,ripoche_determination_1997,liu_direct_2005,chen_single-shot_2007}. At the fastest timescale, the laser pulsewidth $\tau$ is shorter than the fundamental vibrational period and the fastest rotational response time, $\tau \ll \Omega_v^{-1} \ll \Delta t_{rot}$, where $\Omega_v$ is the fundamental vibrational frequency and $\Delta t_{rot} \sim 2T/[j_{max}(j_{max}+1)]$ is the fastest rotational timescale, where $T$ and $j_{max}$ are the rotational revival period and the quantum number of the maximally populated rotational state. Optical laser excitation of vibrational and rotational states can occur by two-photon Raman excitation, but for commonly used $\sim$50-100 fs optical pump pulses, there is insufficient laser bandwidth ($\hbar \Delta\omega <0.04$ eV) to excite vibrational modes in molecules of interest for atmospheric propagation ($\hbar \Omega_v \sim 0.3$ eV for N$_2$), and only rotational states are excited \cite{chen_single-shot_2007}. For example, $\hbar/\Delta t_{rot} \sim 0.03$ eV in N$_2$, where $j_{max} \sim 10$ at room temperature. However, for femtosecond pulse filamentation in gases, significant pulse self-shortening and spectral broadening \emph{can} occur so as to excite vibrational modes even in H$_2$ (in which $\hbar \Omega_v \sim 0.5$ eV), driven by pulse spikes thought to be as short as several femtoseconds \cite{odhner_rovibrational_2009,odhner_filament-driven_2011}. Nevertheless, to our knowledge there has never been a controlled experiment directly measuring the full, time-resolved electronic and rovibrational nonlinear response of a light molecule.

Here, we measure the full absolute electronic and rovibrational nonlinear response of H$_2$ and D$_2$ to intense ultrashort optical pulses up to the ionization threshold of $\sim 10^{14}$ W/cm$^2$. We use the technique of single-shot supercontinuum spectral interferometry (SSSI) \cite{chen_single-shot_2007,kim_single-shot_2002-1}. The 40 fs pump pulse inducing the nonlinear response is sufficiently short to impulsively drive the rotational nonlinearity, but, on its own, cannot drive vibrational modes as discussed above. However, the vibrational component of the nonlinearity contributes to a two-beam coupling phase shift and energy transfer \cite{dogariu_purely_1997,smolorz_femtosecond_2000,wahlstrand_effect_2013}, even for a long pump pulse. We use this phase shift, which is independent of the amplitude of a sufficiently weak probe pulse, to also measure the vibrational component of the optical nonlinearity.

The nonlinear response can be explored using single-beam experiments \cite{nibbering_determination_1997,liu_direct_2005,laban_self-focusing_2010}, but pump-probe techniques allow direct time-resolved observation of the response \cite{chen_single-shot_2007,wahlstrand_effect_2013,wahlstrand_optical_2011-1,wahlstrand_absolute_2012,wahlstrand_high_2012} or the response reconstructed with an auxiliary model for the pump spatio-temporal dependence \cite{ripoche_determination_1997,loriot_measurement_2009,reichert_beam_2015}. 
Single-shot techniques, in particular, eliminate laser pulse-to-pulse fluctuations as a source of error \cite{chen_measurement_2007,chen_single-shot_2007,kim_single-shot_2002-1,wahlstrand_effect_2013,wahlstrand_optical_2011-1,wahlstrand_absolute_2012,wahlstrand_high_2012}. However, it is crucial in pump-probe experiments to properly account for two-beam coupling effects, as we recently showed for plasma and rotational nonlinearities \cite{wahlstrand_effect_2013}.
Two-beam coupling for the case of a vibrational nonlinearity is a primary focus of this paper.

\section{Experiment}

In our SSSI setup, a regenerative Ti:Sapphire amplifier produces 3 mJ, 40 fs pulses centered near 800 nm. Supercontinuum (SC) pulses covering the range 500 nm to 700 nm are generated by weakly focusing a few hundred $\mu$J split from the main pulse into a gas cell filled with approximately 2.5 atm SF$_6$ or Xe. The SC pulse is strongly chirped (see later) and split into probe and reference pulses separated by 2 ps, which are then collinearly combined with an 800 nm pump beam using a dichroic mirror, with the reference pulse earliest in time and the pump and probe pulses temporally overlapping. The collinear pulses propagate through either a thin laser-drilled stainless steel gas flow tube in a vacuum chamber, or through the backfilled chamber without the tube. The probe pulse picks up the wavelength-dependent phase and amplitude change induced in the H$_2$ or D$_2$ gas by the intense pump pulse, which are then extracted from the spectral interference pattern collected in an imaging spectrometer, which also provides transverse 1D spatial resolution ($x$-direction). The absolute nonlinear change in refractive index $\Delta n(x,t) =(k_0 \Leff)^{-1} \Delta \Phi (x,t)$ is then determined, where $k_0$ is the probe central wavenumber, $\Leff$ is the effective gas interaction length, and $\Delta \Phi(x,t)$ is the space and time resolved phase shift extracted from the spectral interferogram. Full details on SSSI and the determination of the absolute index shift are found in references \cite{chen_single-shot_2007,wahlstrand_absolute_2012}.

\begin{figure}
\centering
\includegraphics[width=8cm]{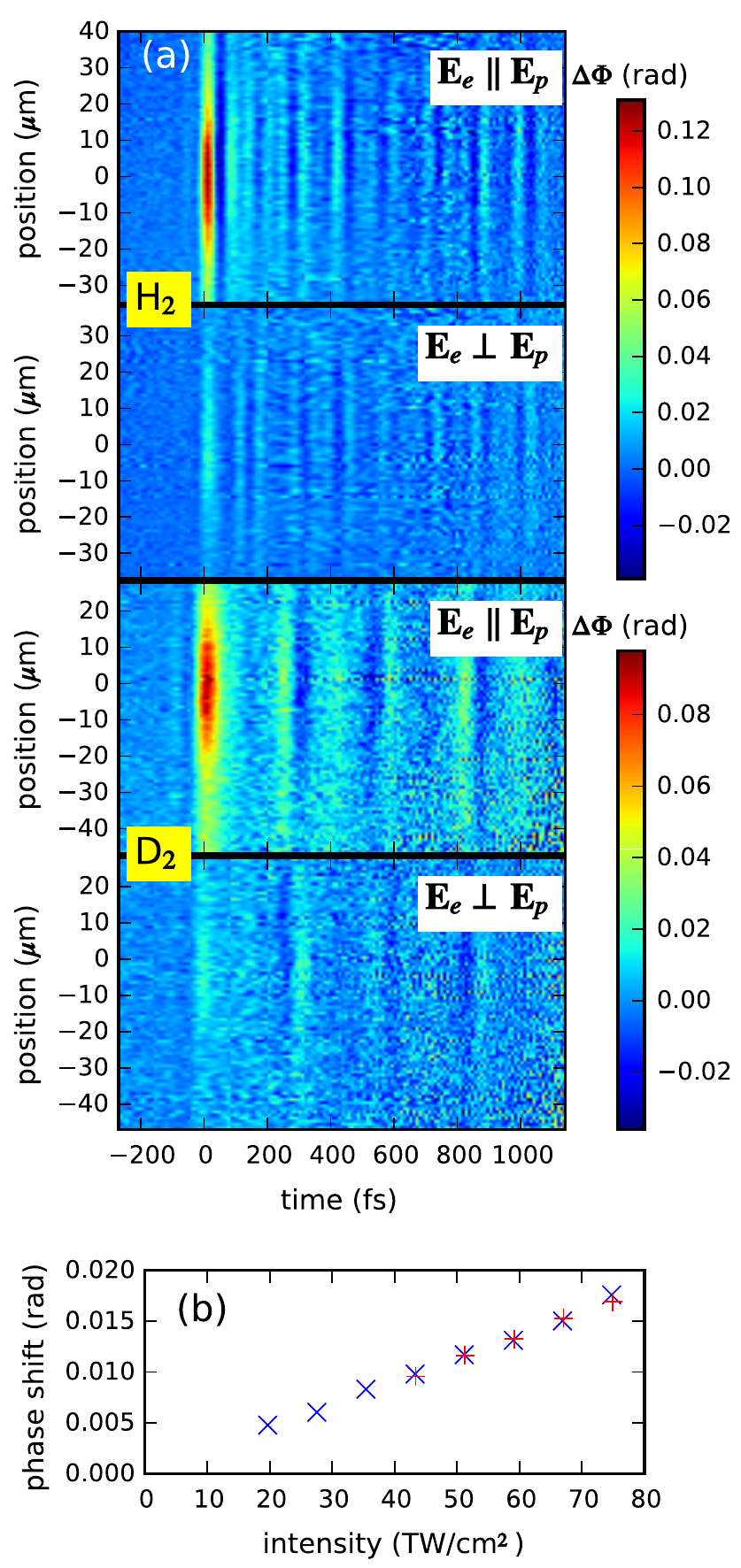}
\caption{(a) Pump-induced 1D space- and time-resolved phase shift $\Delta\Phi(x,t)$ for a 46 TW/cm$^2$, 40 fs pulse in H$_2$ and D$_2$ for the pump $\bEe$ and probe $\bEp$ polarizations parallel and perpendicular. To increase the signal to noise \cite{chen_measurement_2007}, $\Delta\Phi(x,t)$ is extracted from the average of 200 single-shot interferograms. (b) Intensity dependence of the peak phase shift in H$_2$ (+) and D$_2$ ($\times$) extracted from the spatial profile of the response $\Delta \Phi(x,t=0)$ for a pump pulse of peak intensity $\sim 80$ TW/cm$^2$.}
\label{phasemaps}
\end{figure}

An example of the full nonlinear response measured by SSSI is shown in Fig.~\ref{phasemaps}a, which plots the phase shift $\Delta \Phi(x,t)$ measured for H$_2$ and D$_2$ in a backfilled chamber filled to $\sim 0.3$ atm for a peak pump intensity of $\sim 46$ TW/cm$^2$ and pulsewidth 40 fs, where the probe pulse is polarized parallel or perpendicular to the pump. At this intensity the contribution of free electrons from ionization is negligible, without a detectable long timescale negative phase shift on the probe \cite{wahlstrand_optical_2011-1,wahlstrand_high_2012}. As can be seen, there is a prompt response near the center of the pump envelope at $t = 0$, with electronic and vibrational contributions (to be discussed), followed by a modulated delayed response caused by the beating of a coherent superposition of quantized rotational states. For a peak pump intensity of $\sim 80$ TW/cm$^2$ (just below the ionization thresholds of H$_2$ and D$_2$) and a thin gas target to ensure a uniform axial pump intensity \cite{wahlstrand_absolute_2012}, we use the spatially-varying pump intensity imprinted on $\Delta \Phi(x,t=0)$ to plot the phase shift as a function of pump intensity for the probe pulse polarized parallel to the pump pulse in Fig.~\ref{phasemaps}b. As in N$_2$ and the noble gases \cite{wahlstrand_optical_2011-1,wahlstrand_high_2012}, the prompt nonlinear response of H$_2$ or D$_2$ is linear in the intensity, with no saturation in the electronic response observed below the ionization threshold, implying that the nonlinear coefficients measured in this paper apply up to that limit.
As before \cite{wahlstrand_high_2012}, we define the ionization threshold as the peak intensity level at which we observe a long-lived negative probe phase shift above the noise floor.

\section{Rotational response}
\begin{figure}
\centering
\includegraphics[width=8cm]{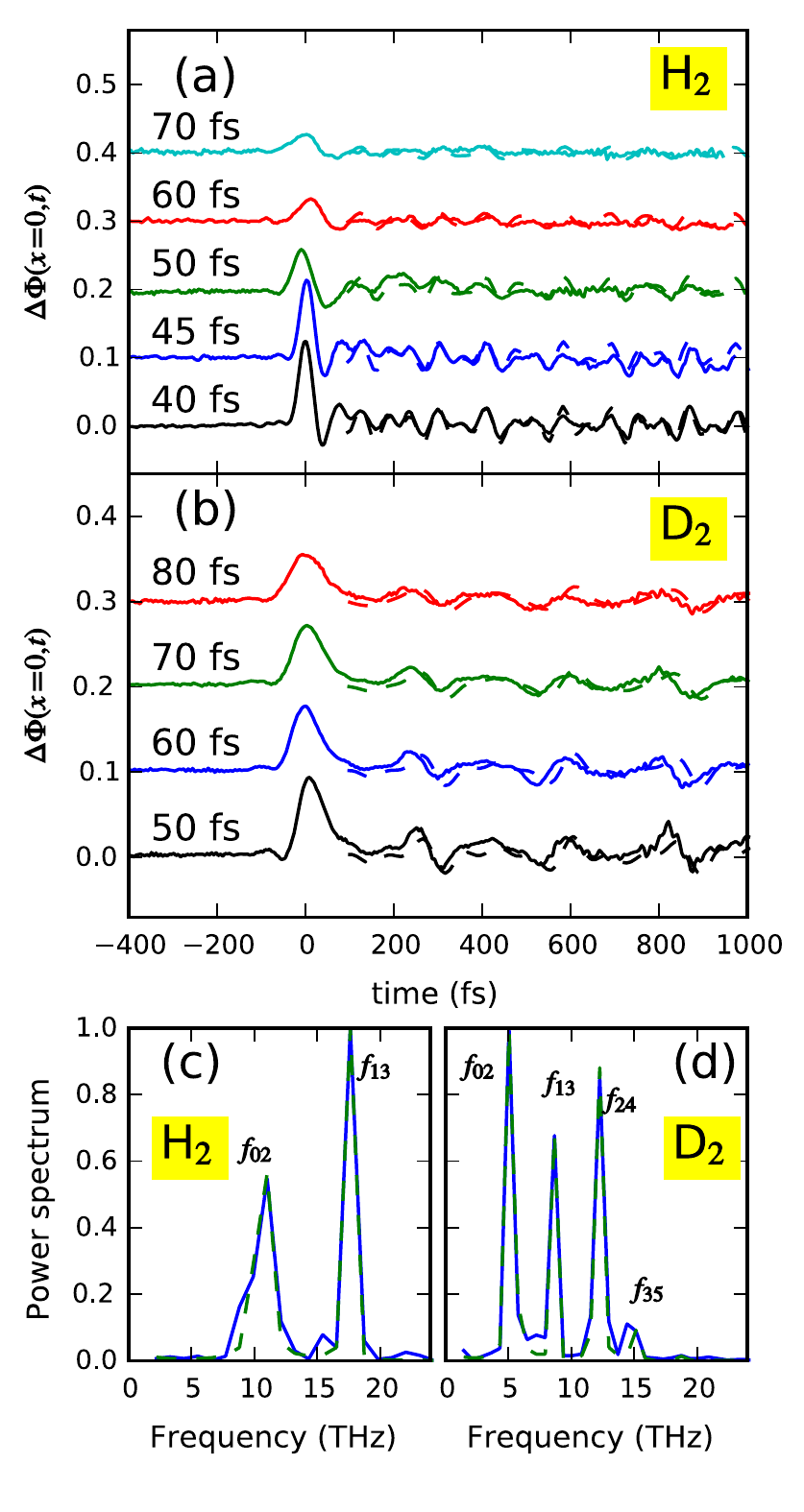}
\caption{Experimental data on the rotational molecular response in H$_2$ and D$_2$. (a,b) Time-dependent phase shift for $\bEe \parallel \bEp$ as a function of pump pulse duration for (a) H$_2$ and (b) D$_2$. Density matrix simulations of the rotational phase response (using fitted values of the rotational constants as described in the text) are shown as dashed lines. (c,d) The discrete Fourier transform (DFT) of the phase shift at times after the 40 fs pump pulse for (c) H$_2$ and (d) D$_2$, showing which rotational coherences are excited. The green dashed curves in (c,d) show the DFT of the fitted density matrix simulations.}
\label{rotational}
\end{figure}

We first discuss the rotational response, as it solely contributes for times long after the pump pulse. In H$_2$ and D$_2$, it depends strongly on the pulse duration. The probe phase shift $\Delta \Phi(x=0,t)$ is shown as a function of pump pulse width in Figs.~\ref{rotational}a and \ref{rotational}b as solid curves, for fixed pump pulse energy of 34 $\mu$J. For efficient excitation of a long-lived rotational coherence, the pump pulse width should be shorter than the characteristic response period. Here, for a Gaussian pulse of duration $\tau$, the efficiency of excitation of the $j \rightarrow k$ rotational coherence is proportional to $e^{-\tau^2 (E_j - E_k)^2/\hbar^2}$ and its contribution to the optical response depends on the initial population in states $j$ and $k$ \cite{chen_single-shot_2007}, where the rotational energies, including the centrifugal correction, are $E_j = hcBj(j+1)-hcDj^2(j+1)^2$, where $j$ is the rotational quantum number, $B$ is the rotational constant, and $D$ is the centrifugal constant. Note the reduced modulation amplitude of the rotational mode beating as the pump pulse duration increases; longer pulses become increasingly less efficient at exciting the rotational coherences. The discrete Fourier transform of $\Delta \Phi(x=0,t)$ for $t > 200$ fs for the shortest pulse width of $\sim 40$ fs is shown in Figs.~\ref{rotational}c and \ref{rotational}d. We observe the $j = 0 \rightarrow 2$ and $1 \rightarrow 3$ transitions in both H$_2$ and D$_2$ and additionally the $j = 2 \rightarrow 4$ and $3 \rightarrow 5$ transitions in D$_2$.

In our determination of the absolute nonlinear response of H$_2$ and D$_2$ to intense fields, it is important to note that \emph{all} parameters needed are provided by our SSSI measurements. For the rotational component of the response, the needed parameters are $B$, $D$ and $\Delta \alpha = \alpha_\parallel - \alpha_\perp$, the molecular polarizability anisotropy. First, the oscillation frequencies measured in Figs.~\ref{rotational}c and \ref{rotational}d yield best fits of $B = 58.9$ ($\pm 3.0$) cm$^{-1}$, $D = 0.05$ ($\pm 0.02$) cm$^{-1}$ for H$_2$, and $B = 29.3$ ($\pm 1.5$) cm$^{-1}$, $D = 0.021$ ($\pm 0.008$) cm$^{-1}$ for D$_2$, in good agreement with the literature values \cite{irikura_experimental_2007,huber_molecular_1979}. The dominant source of uncertainty in the measurements of $B$ and $D$ is from the chirp of the probe pulse, which determines the mapping between frequency and pump-probe delay \cite{kim_single-shot_2002-1}. Quantum mechanical density matrix simulations \cite{chen_single-shot_2007,zahedpour_quantum_2014} of the H$_2$ and D$_2$ rotational response (for $t>100$ fs, after the prompt response) employing these values of $B$ and $D$ are shown in Fig.~\ref{rotational} as dashed lines, indicating excellent agreement.

Determination of $\Delta \alpha$, which scales (as $(\Delta \alpha)^2$) the dependence of the molecular polarizability on the rotational response, requires an absolute measurement of $\Delta n(x,t)$. Here, the experiment of Fig.~\ref{phasemaps} was repeated using a calibrated thin gas target \cite{wahlstrand_absolute_2012}, and the refractive index modulation for $t > 200$ fs was compared to density matrix simulations \cite{chen_single-shot_2007,zahedpour_quantum_2014} of the rotational response to arrive at best fit values of $\Delta \alpha = (3.0 \pm 0.6) \times 10^{-25}$ cm$^3$ for H$_2$ and $\Delta \alpha = (3.0 \pm 0.4)\times 10^{-25}$ cm$^3$ for D$_2$. These values are in agreement with previous scattering-based measurements \cite{bridge_polarization_1966} and calculations \cite{bishop_vibrational_1993}.

\section{Vibrational response}
The rotational contribution to the nonlinearity arises from the increased ensemble polarizability as the initially randomly oriented molecules are torqued into alignment with the optical field \cite{chen_single-shot_2007}. Similarly, a vibrational contribution arises from the optical force pushing the molecule's constituent atoms apart and is expressed as the dependence of the molecular polarizability tensor $\bm{\alpha}$ on changes $Q$ in the internuclear separation, $\bm{\alpha}(Q) = \bm{\alpha}(0)+(\partial \bm{\alpha}/\partial Q)Q$ for $Q$ small compared to the equilibrium internuclear distance of $<1$ \AA.
The vibrational frequencies in H$_2$ and D$_2$ are in the $\sim 100$ THz range, so to observe vibrational excitation directly (as we are able to do with the rotational excitation) one would need a few-cycle optical pulse, as discussed earlier.
For a pulse longer than a few optical cycles, the vibrational response is \emph{adiabatic}.
The laser-induced molecular stretch $Q$, and thus the stretch-induced change in the polarizability, is proportional to the pulse intensity, causing an effectively prompt response. One could in principle infer the size of the vibrational response by observing a reduced nonlinear refractive index for a few-cycle pulse compared with a longer pulse \cite{laban_self-focusing_2010}.
It appears impossible to distinguish the vibrational nonlinearity from the electronic nonlinearity through pumping with a 40 fs pulse alone.
However, we are able to separate them another way. In a pump-probe experiment, interference between the pump and probe beams leads to the generation of a nonlinear refractive index grating, which modifies the nonlinear phase shift of the probe beam. This two-beam coupling phase shift was recently discussed in detail for the bound and free electronic and the rotational components of the optical nonlinearity in diatomic molecules \cite{wahlstrand_effect_2013}.
Here, we use the two-beam coupling phase shift from rovibrational modes to measure the rovibrational component of the optical nonlinearity in H$_2$ and D$_2$.

We model the molecular interaction with a pump plus probe field $\mathbf{E}(\mathbf{r},t) = (1/2)(\hat{\mathbf{e}_e} A_e(t) e^{i(\mathbf{k}_e \cdot \mathbf{r}-\omega_e t)}+\hat{\mathbf{e}_p} A_p(t) e^{i(\mathbf{k}_p \cdot \mathbf{r}-\omega_p t)})+c.c.$, where subscripts $e$ and $p$ refer to pump and probe, $A_e(t)$ and $A_p(t)$ are complex field envelopes (for a weak probe $|A_p/A_e| \ll 1$), and we define $\Delta \omega = \omega_p-\omega_e$.
As described in the Appendix, for an ultrashort optical pulse, the full rovibrational model \cite{demtroder_molecular_2005} can be well approximated by replacing the $j$-dependent vibrational frequencies by a single vibrational mode of frequency $\Omega_v$, so that the total vibrational nonlinear response appears as the polarization $\mathbf{P}^{vib} (t) = N(\partial \alpha/\partial Q) \langle Q \rangle_t \mathbf{E}$, where $\alpha = (1/3) \alpha_\parallel + (2/3) \alpha_\perp$, $N$ is the molecular density, and $\langle Q \rangle_t$ is the time-dependent ensemble-averaged molecular stretch, given by
\begin{equation}
\langle Q \rangle_t \approx \frac{1}{4\mu\OV} \frac{\partial \alpha}{\partial Q} \int_{-\infty}^t \sin[\OV(t'-t)]E^2 dt',
\label{Qsimplified}
\end{equation}
where $\mu$ is the molecular reduced mass.

Inserting the expression for $E$ into Eq.~(\ref{Qsimplified}) and keeping slowly varying terms up to first order in the probe field, we find
\begin{equation}
\langle Q \rangle_t = \langle Q \rangle_t^s + (\langle Q \rangle_t^g e^{i\Delta \bk \cdot \br} + c.c.),
\end{equation}
where $\Delta \bk = \bk_p - \bk_e$, and
\begin{eqnarray}
\langle Q \rangle_t^s &=& \frac{1}{4\mu\Omega_v} \frac{\partial \alpha}{\partial Q} \int_{-\infty}^t \sin[\Omega_v(t'-t)]|A_e(t')|^2 dt', \label{Qsmooth} \\
\langle Q \rangle_t^g &=& \frac{1}{8\mu\Omega_v} \frac{\partial \alpha}{\partial Q} \int_{-\infty}^t \sin[\Omega_v(t'-t)]A_e^*(t') A_p(t') e^{-i\Delta \omega t'} dt'. \label{Qgrating}
\end{eqnarray}
The superscripts denote ``smooth'' and ``grating'' \cite{wahlstrand_effect_2013,wahlstrand_effect_2011}.
The former refers to the ensemble average stretch induced by the pump field only, and the latter to the stretch contribution induced by interference between the pump and probe fields.
These stretch terms are the origin of the vibrational nonlinear response contributing to the polarization oscillating at frequencies near $\omega_p$ (which is measured in SSSI), 
\begin{equation}
\mathbf{P}_p^{vib}(t) = N \frac{\partial \alpha}{\partial Q} [\langle Q \rangle_t^s \bAp(t) e^{i\bk_p \cdot \br-i\omega_p t} + \langle Q \rangle_t^g \bAe(t) e^{i\bk_p \cdot \br - i\omega_e t}].
\end{equation}

For pump pulses of duration longer than the vibrational period $2\pi\Omega_v^{-1} \sim 8$-12 fs in H$_2$ and D$_2$, which applies to virtually all ultrashort pulse optical light sources, the vibrational response is effectively instantaneous. We therefore consider the probe instantaneous nonlinear polarization $\mathbf{P}^{inst}_p (t) = \ceff \mathbf{A}_p (t) e^{i(\mathbf{k}_p \cdot \mathbf{r}-\omega_p t)}$, where the effective nonlinear susceptibility $\ceff$ is related to the nonlinear index shift by $2\pi \ceff = \Delta n = n_2 I_e$, with $n_2$ split into electronic plus vibrational components: $n_2 = n_{2,elec} + n_{2,vib}$, with $n_{2,vib} = n^s_{2,vib} + n^g_{2,vib}$. Note that for a nonresonant electronic response, the smooth and grating terms are the same, so $n_{2,elec} = n_{2,elec}^s + n_{2,elec}^g = 2n_{2,elec}$. The nonlinear refractive index for a weak probe pulse is a factor of two larger than for a pulse acting on itself \cite{wahlstrand_effect_2013}.

To gain immediate physical insight, we first consider the limit of CW pulses.
Then we have for the smooth component, using Eq.~(\ref{Qsmooth}),
\begin{equation}
\langle Q \rangle_t^s = \frac{1}{4\mu \OV} \frac{\partial\alpha}{\partial Q} |A_e|^2 \int_{-\infty}^t \sin[\OV(t'-t)] dt' = \frac{1}{4\mu \OV^2} \frac{\partial\alpha}{\partial Q} |A_e|^2,
\label{smoothngrating}
\end{equation}
and for the grating term, using Eq.~(\ref{Qgrating}),
\begin{eqnarray}
\langle Q \rangle_t^g &\approx& -\frac{1}{4\mu\OV} \frac{\partial \alpha}{\partial Q} A_e^* A_p \int_{-\infty}^t \sin[\OV(t'-t)] e^{-i\Delta \omega t'} dt' \nonumber \\
&\approx& -\frac{1}{4\mu(\Delta \omega^2 - \OV^2)} \frac{\partial \alpha}{\partial Q} A_e^* A_p e^{-i\Delta \omega t}.
\end{eqnarray}
These expressions then yield
\begin{eqnarray}
n_{2,vib}^s &=& \frac{4\pi^2 N}{n_0 c \mu \Omega_v^2} \left(\frac{\partial \alpha}{\partial Q}\right)^2, \label{vibsmooth} \\
n_{2,vib}^g &=& \frac{4\pi^2 N}{n_0 c \mu (\Omega_v^2-\Delta\omega^2)} \left(\frac{\partial \alpha}{\partial Q}\right)^2.
\end{eqnarray}

Note that the vibrational nonlinear response picked up by the probe is greatly enhanced by the grating term $n_{2,vib}^g$ for $\Delta \omega \sim \Omega_v$, and we exploit the dependence on $\Delta \omega/\Omega_v$ of this resonant two-beam coupling effect to measure the vibrational component of the refractive index. We note that in continuous wave experiments, it is well known that the vibrational nonlinearity depends strongly on such resonant coupling \cite{lundeen_nonresonant_1983,shelton_measurements_1994}. Here we use the effect in the ultrafast domain to separate the electronic and vibrational contributions to the prompt nonlinearity.
The expression above for $n_{2,vib}^g$ is singular when $\Delta \omega = \OV$ because of the assumption of infinitely long pulses.
For a long, finite pump pulse of the form $A_e(t) = A_{e0} e^{-t^2/(2t_e^2)}$ (where $t_e \gg \OV^{-1}$), it is straightforward to derive a well-behaved expression for the grating vibrational nonlinearity by integrating Eq.~(\ref{Qsmooth}),
\begin{equation}
n_{2,vib}^g = \frac{4\pi^2 N t_e}{n_0 c \mu \Omega_v} \left[ F(t_e(\Delta\omega-\Omega_v)) - F(t_e(\Delta\omega+\Omega_v)) \right] \left(\frac{\partial \alpha}{\partial Q}\right)^2,
\label{dawson}
\end{equation}
where $F(x)$ is the Dawson function.

\begin{figure}
\centering
\includegraphics[width=8cm]{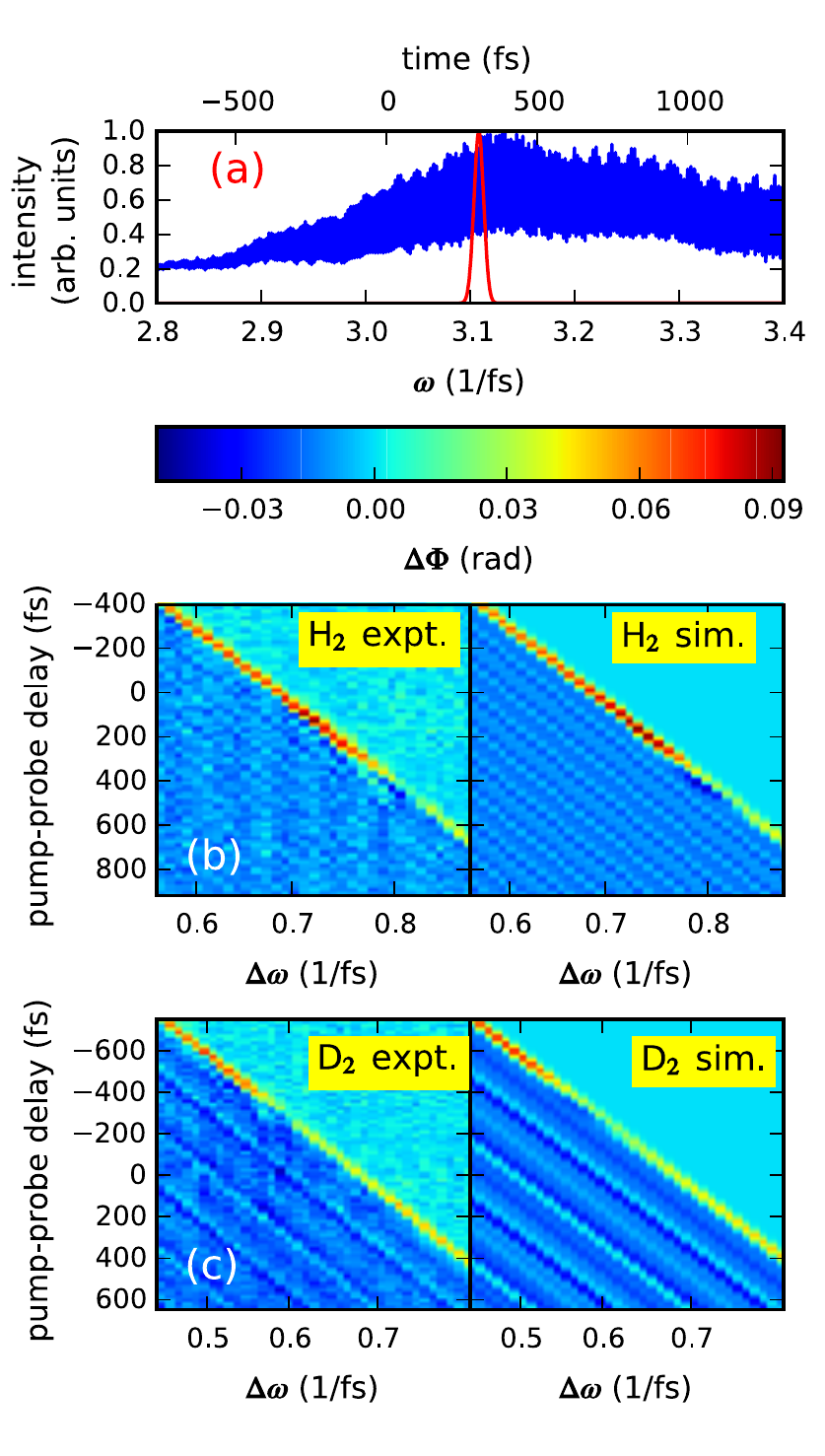}
\caption{Phase shift of chirped SC probe vs.~$\Delta \omega = \omega_{p,overlap} - \omega_e$. The measured probe spectral phase is $\phi(\omega) = \beta_2(\omega-\omega_0)^2 + \beta_3(\omega-\omega_0)^3$, with $\beta_2 = 1675$ fs$^2$, $\beta_3 = 396$ fs$^3$, and $\omega_0 = 3.05$ fs$^{-1}$. The time delay of the SC frequency component at $\omega$ (left axes) is given by $\phi'(\omega)$. (a) Probe/reference SC interference spectrum with variably delayed pump pulse superposed. The top axis shows the time corresponding to each frequency component. (b) Experimental and best-fit rovibrational simulation plots for (b) H$_2$ and (c) D$_2$. The best fit simulations enabled extraction of $n_{2,elec}$ and $\partial \alpha/\partial Q$, shown in Table I. }
\label{vibrational_phasemaps}
\end{figure}

A very useful representation of our resonant two-beam coupling results is a 2D plot of SC probe spectral phase shift vs. $\Delta \omega = \omega_{p,overlap} - \omega_e$, where $\omega_{p,overlap}$ is the probe frequency which overlaps in time with the pump pulse, as shown schematically in Fig.~\ref{vibrational_phasemaps}a. Measurements and simulations of SC probe phase shift vs.~$\Delta \omega$ in H$_2$ are shown in Fig.~\ref{vibrational_phasemaps}b and for D$_2$ in Fig.~\ref{vibrational_phasemaps}c. The time delay of the SC frequency component at $\omega$ (left axes) is given by $\phi'(\omega)$, where $\phi(\omega)$ is the chirped SC spectral phase (see Fig.~\ref{vibrational_phasemaps} caption). For each value of $\Delta \omega$, scanning top to bottom in the plots (negative to positive delay) shows the SC phase shift abruptly changing as the molecular nonlinearity is excited by the arriving pump pulse. The initial perturbation is dominated by the electronic and vibrational response, followed at longer times by rotational revivals. Near $\Delta \omega \sim 0.78$ fs$^{-1}$ in H$_2$ ($\lambda_{p,overlap}$ = 600 nm) and $\Delta \omega \sim 0.55$ fs$^{-1}$ in D$_2$ ($\lambda_{p,overlap}$ = 645 nm), the phase changes strongly, consistent with resonant two-pulse coupling with molecular resonances at $\Omega_v$ = 0.78 fs$^{-1}$ for H$_2$ and 0.55 fs$^{-1}$ for D$_2$ \cite{huber_molecular_1979}. 

\section{Extraction of nonlinear coefficients}

We previously found the rotational response parameters $\Delta \alpha$, $B$, and $D$ from the response at time delays following the pump pulse.
Fitting the one-mode vibrational simulation to the measurements of Fig.~\ref{vibrational_phasemaps} now enables extraction of the two remaining nonlinear coefficients, $n_{2,elec}$ and $\partial \alpha/\partial Q$. The best fit coefficients are listed in Table I, with their associated simulation results shown in the right-hand panels of Figs.~\ref{vibrational_phasemaps}b and \ref{vibrational_phasemaps}c. The $n_{2,elec}$ and $n_{2,vib}$ values we find are consistent with previous measurements using long, low intensity pulses based on harmonic generation \cite{shelton_nonlinear-optical_1990}, the optical Kerr effect \cite{rosasco_dispersion_1986}, and coherent anti-Stokes Raman scattering \cite{hahn_measurement_1995}, and the values of $\partial \alpha/\partial Q$ are consistent with previous calculations \cite{bishop_vibrational_1993}. The most important source of error is the determination of the pump laser intensity, which relies on accurate measurement of the spatial pump beam profile at the gas target.

\begin{table*}
\centering
\caption{Measured electronic Kerr coefficient $n^s_{2,elec}$ at 1 atm (at probe wavelength of 600 nm and pump wavelength of 800 nm), polarizability anisotropy $\Delta \alpha$, and polarizability derivative $\partial \alpha/\partial Q$. Comparison is made with previous experimental and theoretical work. The rightmost column shows $n_2$ estimates at 1 atm based on our experimental results for a long pulse at 800 nm.}
\begin{tabular}{ccccccccccc}
\hline
Gas & \multicolumn{2}{c}{$n_{2,elec}^s$ ($10^{-20}$ cm$^2$/W)} & \multicolumn{2}{c}{$\Delta \alpha$ ($10^{-25}$ cm$^3$)} & \multicolumn{2}{c}{$\partial \alpha/\partial Q$ ($10^{-16}	$ cm$^2$)} & \multicolumn{3}{c}{$n_2$ at 800 nm ($10^{-20}$ cm$^2$/W)}\\
\hline
 & This expt. & Ref.~\cite{shelton_measurements_1994} & This expt. & Ref.~\cite{bridge_polarization_1966} & This expt. & Ref.~\cite{bishop_vibrational_1993} & $n_{2,rot}$ & $n_{2,vib}$ & $n_{2,tot}$ \\
 \hline
H$_2$ & $6.5 \pm 1.0$ & 6.37 & $3.0 \pm 0.6$ & 3.14 & $1.3\pm 0.2$ & 1.24 & 2.7 & 1.0 & 10.2 \\
D$_2$ & $5.8 \pm 1.3$ & 6.20 & $3.0 \pm 0.4$ & 2.99 & $1.4 \pm 0.3$ & 1.24 & 3.9 & 1.2 & 10.9 \\
\hline
\end{tabular}
  \label{tab:shape-functions}
\end{table*}

\begin{figure}
\centering
\includegraphics[width=8cm]{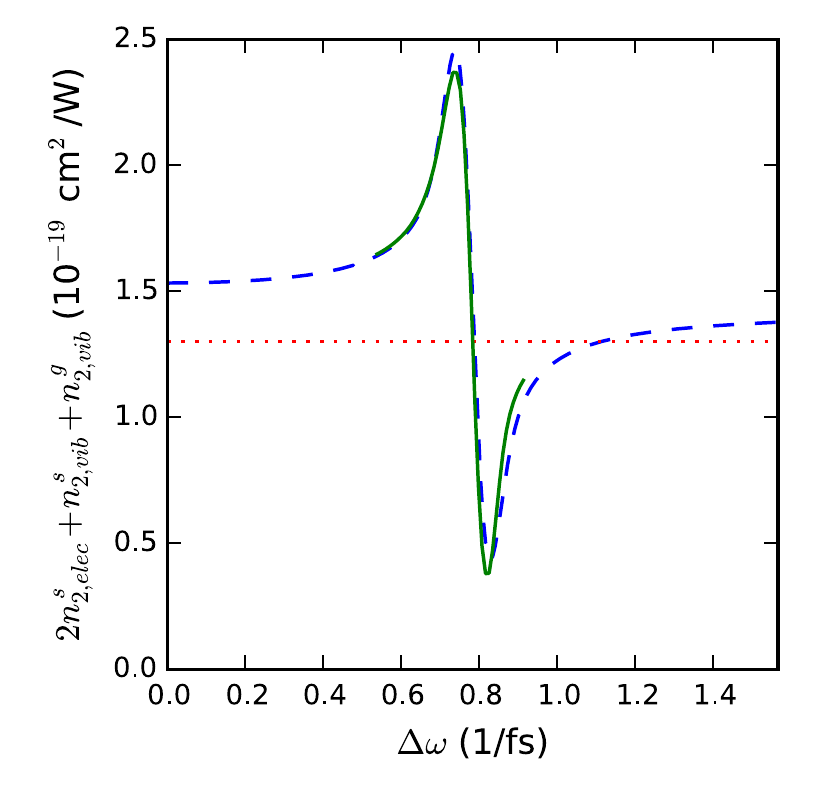}
\caption{
Calculated prompt nonlinear index $2n^s_{2,elec} + n_{2,vib}^s + n_{2,vib}^g$ of H$_2$ experienced by the probe as a function of the pump-probe beat frequency $\Delta\omega$ for a 40 fs pulse centered at 800 nm, calculated using Eq.~(\ref{dawson}) (dashed blue) and numerically (solid green). The electronic component $2n^s_{2,elec}$ is shown as a red dotted line for comparison.}
\label{simple_vib}
\end{figure}

The total prompt (non-rotational) nonlinear refractive index $2n^s_{2,elec} + n_{2,vib}^s + n_{2,vib}^g$ experienced by the probe in H$_2$ is plotted as a function of $\Delta \omega$ in Fig.~\ref{simple_vib}a, as calculated for a 40 fs pump pulse and a broad chirped SC probe, both using the analytical expression [Eq.~(\ref{dawson})] and calculated numerically. The electronic and vibrational nonlinear coefficients are taken from Table I. The electronic component of the nonlinear index $2n^s_{2,elec}$ is shown as a red dashed line for comparison.

\begin{figure}
\centering
\includegraphics[width=8cm]{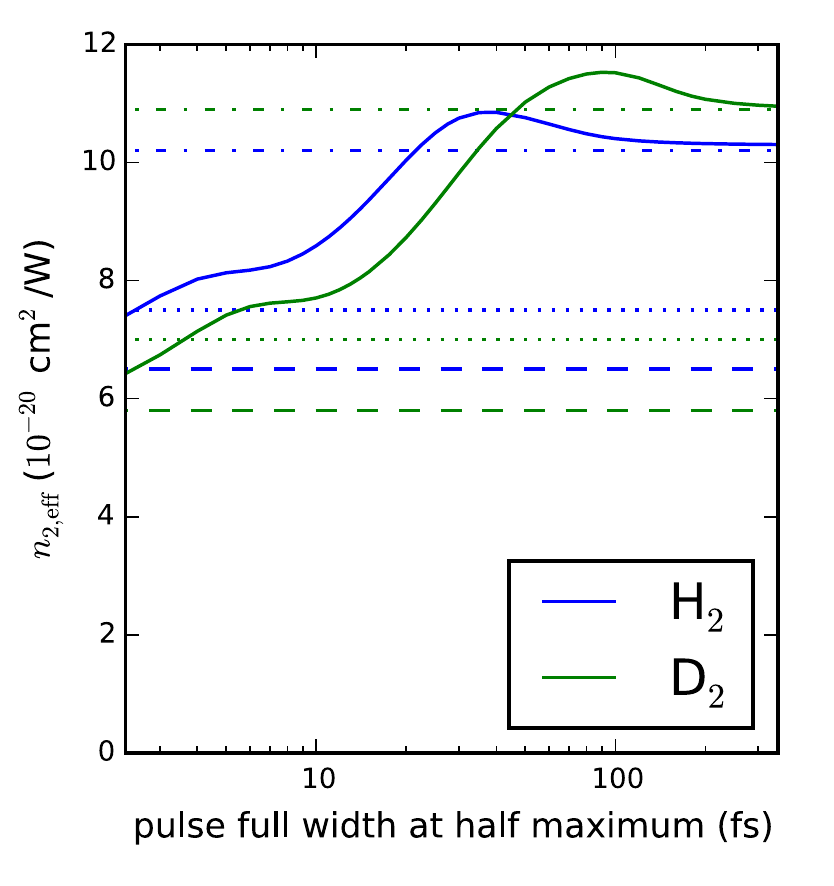}
\caption{
Pulse width dependent effective $n_2$ in H$_2$ (solid blue) and D$_2$ (solid green) calculated from the experimental results. Horizontal dashed lines show $n^s_{2,elec}$, dotted lines show $n^s_{2,elec}+n_{2,vib}$, and dash-dotted lines show $n^s_{2,elec}+n_{2,vib}+n_{2,rot}$ for H$_2$ (blue) and D$_2$ (green).}
\label{pulse_width_dep}
\end{figure}

As described earlier, the nature of the rotational and vibrational response depends on the pulse width; pulses that are too long to impulsively excite rotational or vibrational modes experience an effective adiabatic response.
To illustrate this, we calculate an intensity-weighted effective nonlinear coefficient for a pulse of the form $I(t) = I_0 g(t)$, where $g(t) = \exp(-t^2/t_e^2)$,
\begin{equation}
n_{2,\mathrm{eff}} = \frac{\lambda}{2\pi L I_0} \frac{\int_{-\infty}^{\infty} \Delta \Phi(t) g(t) dt}{\int_{-\infty}^{\infty} g(t) dt},
\end{equation}
where $\lambda$ is the laser wavelength, $I(t)$ is the pulse intensity envelope, and $\Delta \Phi(t)$ is the calculated time-dependent phase shift over an interaction length $L$.
The effective Kerr coefficient $n_{2,\mathrm{eff}}$ is plotted as a function of pulse full width at half maximum in H$_2$ and D$_2$ in Fig.~\ref{pulse_width_dep}.

The electronic component $n^s_{2,elec}$ in each gas is shown as dashed lines.
It can be seen that even at the shortest pulse durations simulated, the effective nonlinearity is larger than $n^s_{2,elec}$ because of the vibrational contribution.
Near 8 fs, $n_{2,\mathrm{eff}}$ is approximately $n^s_{2,elec} + n_{2,vib}$ (shown as dotted lines).
For longer pulses, the rotational response contributes, and the effective nonlinearity eventually levels off at $n^s_{2,elec} + n_{2,vib}+n_{2,rot}$ (shown as dash-dotted lines).

An expression for the rotational adiabatic coefficient $n_{2,rot}$, analogous to the vibrational expression given here [Eq.~(\ref{vibsmooth})], is given in Ref.~\cite{wahlstrand_absolute_2012}.
Values of $n_{2,rot}$ and $n_{2,vib}$ and the total long pulse nonlinear coefficient $n_{2,tot} = n_{2,elec}^s+n_{2,rot} + n_{2,vib}$ are given in the rightmost columns of Table I.
The long pulse coefficients $n_{2,rot}$ and $n_{2,vib}$ apply to pulses of duration greater than $1/f$, where $f$ is the lowest frequency vibrational or rotational mode.
Given the rotational mode spectrum shown in Figs.~\ref{rotational}c and \ref{rotational}d, the pulse duration above which the rotational response can be considered adiabatic is $\sim 100$ fs in H$_2$ and $\sim 200$ fs in D$_2$, in agreement with the results in Fig.~\ref{pulse_width_dep}.
Given the vibrational mode frequencies, the vibrational adiabatic coefficient $n_{2,vib}$ should apply to pulses longer than approximately $\sim 8$ fs in H$_2$ and $\sim 12$ fs in D$_2$.
These arguments support the results in Fig.~\ref{pulse_width_dep}.

\section{Conclusions}
In summary, we have demonstrated single shot measurement of the full absolute electronic and rovibrational nonlinear response of H$_2$ and D$_2$ to intense ultrashort optical pulses up to the ionization threshold of $\sim 10^{14}$ W/cm$^2$. The presence of the electronic, rotational, and vibrational contributions all in one set of 2D experimental traces enables the determination of the relative contribution of each to the nonlinear response.
Importantly, one does not need a few femtosecond pump pulse to excite the vibrational nonlinearity; our pump-probe configuration promotes its excitation through two-beam coupling, even for much longer pump pulses. The nonlinear coefficients measured are applicable over a very wide range of laser pulse widths and intensities below the ionization threshold, and because they apply to the very simplest molecules they constitue a fundamental benchmark for theory and simulation.

\section*{Acknowledgments}
The authors thank Y.-H. Chen for early discussions.
This research was supported by the National Science Foundation, the Air Force Office of Scientific Research, and the Army Research Office.

\appendix

\section{Rovibrational model}
We consider a nonrigid rotor system for a diatomic molecule where the atoms are assumed to be bound by a spring with a natural frequency $\OV$.
This assumption is valid as long as only the lowest two vibrational states are involved, which is a very good approximation for describing the experimental conditions.
We define $Q$ as the change in atomic separation with respect to equilibrium $R_E$ and $\theta$ and $\phi$ as polar and azimuthal angles, respectively.
Eigenfunctions of the nonrigid rotor are \cite{demtroder_molecular_2005}
\begin{math}
\psi_{njm} (Q,\theta,\phi) = F_{nj} (Q) Y_j^m (\theta,\phi),
\end{math}
where
\begin{multline}
F_{nj}(Q) = \frac{1}{2^n n!} \left(\frac{2\mu\OV}{\pi\hbar}\right)^{1/4} \exp \left[-\frac{\mu\OV}{\hbar} \left(Q-\frac{D}{B} R_E j(j+1) \right)^2 \right] \\ \times H_n \left[ \left( \frac{2\mu\OV}{\hbar} \right)^{1/2} \left( Q-\frac{D}{B} R_E j(j+1) \right)^2 \right],
\end{multline}
$Y_j^m(\theta,\phi)$ is a spherical harmonic, and $H_n(R)$ is a Hermite-Gaussian function.
Here, $n$ is the vibrational quantum number, $j$ is the total angular momentum quantum number, and $m$ is the quantum number for angular momentum along the $z$ direction.
The corresponding energies are
\begin{equation}
\hbar \omega_{nj} = \hbar \OV \left(n+\frac{1}{2} \right) +\frac{\hbar^2 j(j+1)}{\mu R_E^2} - \frac{\hbar^4 j^2(j+1)^2}{\OV^2 \mu R_E^6}.
\end{equation}

We treat the optical field $E$ as a perturbation, using the potential
\begin{equation}
V = -\frac{1}{2} E^2 (\alpha_{\perp} +\Delta\alpha \cos^2 \theta) -\frac{1}{2} E^2 \frac{\partial \alpha_\perp}{\partial Q} Q -\frac{1}{2} E^2 \frac{\partial \Delta\alpha_\perp}{\partial Q} Q \cos^2 \theta.
\label{potential}
\end{equation}
Here $\alpha_\perp$ is the polarizability perpendicular to the optical axis and $\Delta\alpha$ is the polarizability anisotropy, both for the lowest vibrational state (we assume that the molecular polarizability is independent of rotational state).

To find the nonlinear optical response for the probe field parallel to the pump field, we need to calculate the polarization from the time-dependent, ensemble averaged induced dipole moment along the optical field direction,
\begin{equation}
P(t) = N \langle p \rangle_t = \left( \alpha_\perp +\Da \langle \cos^2 \theta \rangle_t + \frac{\partial \alpha_\perp}{\partial Q} \langle Q \rangle_t + \frac{\partial \Da}{\partial Q} \langle Q \cos^2 \theta \rangle_t \right) E(t)
\label{totalpol}
\end{equation}
We find the expressions in angle brackets on the right hand side from the density matrix $\rho$, assuming a thermally populated initial state.
It is sufficient to calculate the first-order perturbation solution $\rho^1$ in $V$, which satisfies $\partial \rho^1_{njmn'j'm'}/\partial t = -i(\omega_{nj}-\omega_{n'j'}) \rho^1_{njmn'j'm'} +(i/\hbar) [\rho^0,V]_{njmn'j'm'}$.

\subsection{Rotational}
This has been discussed in detail in previous papers \cite{chen_single-shot_2007,wahlstrand_effect_2013} 
and except for the centrifugal change in rotational energies, it is unaffected by the nonrigidness of the molecule, so we just give the solution,
\begin{equation}
\langle \cos^2 \theta \rangle_t = \frac{1}{3} + \frac{2}{15} \frac{\Da}{\hbar} \sum_{nj} \frac{j(j-1)}{2j-1}\left( \frac{\rho^0_{n,j,n,j}}{2j+1} - \frac{\rho^0_{n,j-2,n,j-2}}{2j-3} \right) \int_{-\infty}^t \sin[\omega_{n,j,n,j-2} (t'-t) E^2(t') dt',
\end{equation}
where $\omega_{n,j,n',j'} = \omega_{nj}-\omega_{n'j'}$ and $\rho^0_{n,j,n,j}$ is the initial population in state ($n$,$j$).

\subsection{Rovibrational}
Using
\begin{equation}
\langle n | Q | n' \rangle = \left( \frac{\hbar}{4\mu \OV} \right)^{1/2}  [(n'+1)\delta_{n,n'+1} + n^{\prime 1/2}\delta_{n,n'-1}],
\label{vibexpand}
\end{equation}
the purely vibrational term in the potential, in the basis of eigenfunctions, is
\begin{equation}
V_{nn'} = -\frac{1}{2} \frac{\partial \alpha_\perp}{\partial Q} \langle n | Q | n' \rangle E^2 = -\frac{1}{2} \frac{\partial \alpha_\perp}{\partial Q} \left( \frac{\hbar}{4\mu \OV}\right)^{1/2} [(n'+1)\delta_{n,n'+1} + n^{\prime 1/2}\delta_{n,n'-1}] E^2.
\end{equation}
The nonzero first-order perturbation solutions for this term in the potential are
\begin{eqnarray}
\rho^1_{n,n-1} &=& -\frac{i}{2\hbar} \frac{\partial \alpha_\perp}{\partial Q} \left( \frac{\hbar}{4\mu \OV}\right)^{1/2} n^{1/2} (\rho^0_{n,n} - \rho^0_{n-1,n-1}) \int_{-\infty}^t \exp[i\omega_{n,n-1} t'] E^2(t') dt' \\
\rho^1_{n,n+1} &=& -\frac{i}{2\hbar} \frac{\partial \alpha_\perp}{\partial Q} \left( \frac{\hbar}{4\mu \OV}\right)^{1/2} (n+1)^{1/2} (\rho^0_{n,n} - \rho^0_{n+1,n+1}) \int_{-\infty}^t \exp[i\omega_{n,n+1} t'] E^2(t') dt'.
\end{eqnarray}

In the basis of eigenfunctions, the rotational-vibrational term in the potential is
\begin{eqnarray}
V_{njmn'j'n'} &=& -\frac{1}{2} \frac{\partial \Da}{\partial Q} \langle n | Q \cos^2 \theta | n' \rangle E^2 \\
&=& -\frac{1}{2} \frac{\partial \Da}{\partial Q} \left( \frac{\hbar}{4\mu \OV}\right)^{1/2} [(n'+1)\delta_{n,n'+1} + n^{\prime 1/2}\delta_{n,n'-1}] \nonumber \\ &&
\times\left[ B_+^m (j') \delta_{j,j'+2} + B_-^m (j') \delta_{j,j'-2} + \left(\frac{1}{3} +B_0^m(j') \right) \delta_{jj'}\right] E^2,
\end{eqnarray}
where $B_+^m(j)$, $B_-^m(j)$, and $B_0^m(j)$ are given in \cite{wahlstrand_effect_2013}.
The nonzero first-order perturbation solutions for this term in the potential are
\begin{eqnarray}
\rho_{n,j,n-1,j} &=& K n^{1/2} \left( \frac{1}{3} +B_0^m(j) \right) \left(\rho^0_{n,j,n,j} - \rho^0_{n-1,j,n-1,j} \right) \int_{-\infty}^t e^{i\omega_{n,j,n-1,j} (t'-t)} E^2(t') dt' \nonumber \\
\rho_{n,j,n+1,j} &=& K (n+1)^{1/2} \left( \frac{1}{3} +B_0^m(j) \right) \left(\rho^0_{n,j,n,j} - \rho^0_{n+1,j,n+1,j} \right) \int_{-\infty}^t e^{i\omega_{n,j,n+1,j} (t'-t)} E^2(t') dt' \nonumber \\
\rho_{n,j,n-1,j-2} &=& K n^{1/2} B_+^m (j-2) \left(\rho^0_{n,j,n,j} - \rho^0_{n-1,j-2,n-1,j-2} \right) \int_{-\infty}^t e^{i\omega_{n,j,n-1,j-2} (t'-t)} E^2(t') dt' \nonumber \\
\rho_{n,j,n+1,j-2} &=& K (n+1)^{1/2} B_+^m (j-2)\left(\rho^0_{n,j,n,j} - \rho^0_{n+1,j,n+1,j} \right) \int_{-\infty}^t e^{i\omega_{n,j,n+1,j-2} (t'-t)} E^2(t') dt'\nonumber \\
\rho_{n,j,n-1,j+2} &=& K n^{1/2} B_-^m (j+2)\left(\rho^0_{n,j,n,j} - \rho^0_{n-1,j+2,n-1,j+2} \right) \int_{-\infty}^t e^{i\omega_{n,j,n-1,j+2} (t'-t)} E^2(t') dt'\nonumber \\
\rho_{n,j,n+1,j+2} &=& K (n+1)^{1/2} B_-^m (j+2)\left(\rho^0_{n,j,n,j} - \rho^0_{n+1,j+2,n+1,j+2} \right) \int_{-\infty}^t e^{i\omega_{n,j,n+1,j+2} (t'-t)} E^2(t') dt'\nonumber 
\end{eqnarray}
where $K=-i\hbar^{1/2}/[2\hbar(4\mu \OV)^{1/2}] (\partial \Da/\partial Q)$.

Adding all these terms and using $\langle Q \rangle_t  =$ Tr$[\rho^1(t) Q]$,
\begin{equation}
\langle Q \rangle_t = \sum_{n,j}\frac{n}{4\mu\OV} \left(\frac{\partial \alpha_\perp }{\partial Q} + \frac{1}{3} \frac{\partial \Delta \alpha}{\partial Q} \right) (\rho^0_{n,j,n,j}-\rho^0_{n-1,j,n-1,j} ) \int_{-\infty}^t \sin[\omega_{n,j,n-1,j}(t'-t)]E^2 dt'.
\end{equation}
Further assuming that only the lowest vibrational state is initially populated, i.e. $\rho^0_{njnj} = 0$ for all $n>0$, we find
\begin{equation}
\langle Q \rangle_t = -\sum_{j}\frac{1}{4\mu\OV} \left(\frac{\partial \alpha_\perp }{\partial Q} + \frac{1}{3} \frac{\partial \Delta \alpha}{\partial Q} \right) \rho^0_{0,j,0,j}  \int_{-\infty}^t \sin[\omega_{1,j,0,j}(t'-t)]E^2 dt'.
\end{equation}
We can express this in terms of an orientationally averaged polarizability derivative
\begin{equation}
\frac{\partial \alpha}{\partial Q} = \frac{\partial \alpha_\perp }{\partial Q} + \frac{1}{3} \frac{\partial \Delta \alpha}{\partial R},
\end{equation}
as
\begin{equation}
\langle Q \rangle_t = -\sum_{j}\frac{1}{4M\OV} \frac{\partial \alpha}{\partial Q} \rho^0_{0,j,0,j}  \int_{-\infty}^t \sin[\omega_{1,j,0,j}(t'-t)]E^2 dt'.
\label{expect1c}
\end{equation}

Using $\langle Q \cos^2 \theta \rangle_t  =$ Tr$[\rho^1(t) Q \cos^2 \theta]$, we find
\begin{multline}
\langle Q \cos^2 \theta \rangle_t = \frac{1}{4\mu\OV} \sum_{n,j} \frac{n}{3} \left[ \frac{\partial \alpha_\perp }{\partial Q} + \right. \\ \left. \frac{1}{3} \left( 1+\frac{4}{5} \frac{j+1}{(2j-1)(2j+3)} \right) \frac{\partial \Delta \alpha}{\partial Q} \right] (\rho^0_{n,j,n,j} - \rho^0_{n-1,j,n-1,j}) \int_{-\infty}^t \sin[\omega_{n,j,n-1,j}(t'-t)]E^2 dt'  \\
+ \frac{1}{4\mu\OV} \sum_{n,j} \frac{2}{15} \frac{\partial \Delta \alpha}{\partial Q} \frac{j(j-1)}{2j-1} \left[n \left( \frac{\rho^0_{n,j,n,j}}{2j+1} -\frac{\rho^0_{n-1,j-2,n-1,j-2}}{2j-3} \right)\int_{-\infty}^t \sin[\omega_{n,j,n-1,j-2}(t'-t)]E^2 dt' \right. \\ \left. +(n+1) \left( \frac{\rho^0_{n,j,n,j}}{2j+1} -\frac{\rho^0_{n+1,j-2,n+1,j-2}}{2j-3} \right) \int_{-\infty}^t \sin[\omega_{n,j,n+1,j-2}(t'-t)]E^2 dt'\right].
\end{multline}

Restricting the initial population to $n=0$, we have
\begin{multline}
\langle Q \cos^2 \theta \rangle = \frac{1}{4\mu\OV} \sum_{j} \frac{1}{3} \left[ \frac{\partial \alpha_\perp }{\partial Q} + \right. \\ \left. \frac{1}{3} \left( 1+\frac{4}{5} \frac{j+1}{(2j-1)(2j+3)} \right) \frac{\partial \Delta \alpha}{\partial Q}\right](- \rho^0_{0,j,0,j}) \int_{-\infty}^t \sin[\omega_{1,j,0,j}(t'-t)]E^2 dt' \\
+ \frac{1}{4\mu\OV} \sum_{j} \frac{2}{15} \frac{\partial \Delta \alpha}{\partial Q} \frac{j(j-1)}{2j-1} \left[ -\frac{\rho^0_{0,j-2,0,j-2}}{2j-3}\int_{-\infty}^t \sin[\omega_{1,j,0,j-2}(t'-t)]E^2 dt' \right. \\ \left. +  \frac{\rho^0_{0,j,0,j}}{2j+1} \int_{-\infty}^t \sin[\omega_{0,j,1,j-2}(t'-t)]E^2 dt'\right].
\end{multline}
Using the orientationally averaged polarizability derivative as before,
\begin{multline}
\langle Q \cos^2 \theta \rangle = -\frac{1}{4\mu\OV} \sum_{j} \frac{1}{3} \left[ \frac{\partial \alpha }{\partial Q} + \left(\frac{4}{15} \frac{j+1}{(2j-1)(2j+3)} \right) \frac{\partial \Delta \alpha}{\partial Q} \right] \rho^0_{0,j,0,j} \int_{-\infty}^t \sin[\omega_{1,j,0,j}(t'-t)]E^2 dt'  \\
+ \frac{1}{4\mu\OV} \sum_{j} \frac{2}{15} \frac{\partial \Delta \alpha}{\partial Q} \frac{j(j-1)}{2j-1} \left[ \frac{\rho^0_{0,j-2,0,j-2}}{2j-3}\int_{-\infty}^t \sin[\omega_{1,j,0,j-2}(t'-t)]E^2 dt' \right. \\ \left. -  \frac{\rho^0_{0,j,0,j}}{2j+1} \int_{-\infty}^t \sin[\omega_{0,j,1,j-2}(t'-t)]E^2 dt'\right].
\label{expect2c}
\end{multline}

\subsection{Total rovibrational model}
Combining Eqs.~(\ref{expect1c}) and (\ref{expect2c}), the rovibrational terms in Eq.~(\ref{totalpol}) are
\begin{multline}
\langle Q \rangle \frac{\partial \alpha_\perp }{\partial Q} + \langle Q \cos^2 \theta \rangle \frac{\partial \Delta \alpha }{\partial Q} =  -\frac{1}{4\mu\OV} \left\{ \frac{\partial \alpha_\perp }{\partial Q} \sum_{j}\frac{\partial \alpha}{\partial Q} \rho^0_{0,j,0,j}  \int_{-\infty}^t \sin[\omega_{1,j,0,j}(t'-t)]E^2 dt'  \right. \\
\left. + \frac{\partial \Delta \alpha}{\partial Q} \sum_{j} \frac{1}{3} \left[ \frac{\partial \alpha }{\partial Q} + \left(\frac{4}{15} \frac{j+1}{(2j-1)(2j+3)} \right) \frac{\partial \Delta \alpha}{\partial Q} \right] \rho^0_{0,j,0,j} \int_{-\infty}^t \sin[\omega_{1,j,0,j}(t'-t)]E^2 dt' \right. \\
\left. + \frac{\partial \Delta \alpha}{\partial Q} \sum_{j} \frac{2}{15} \frac{\partial \Delta \alpha}{\partial Q} \frac{j(j-1)}{2j-1} \left[ \frac{\rho^0_{0,j-2,0,j-2}}{2j-3}\int_{-\infty}^t \sin[\omega_{1,j,0,j-2}(t'-t)]E^2 dt' \right. \right. \\ \left. \left. -  \frac{\rho^0_{0,j,0,j}}{2j+1} \int_{-\infty}^t \sin[\omega_{0,j,1,j-2}(t'-t)]E^2 dt'\right] \right\},
\end{multline}
and we can group terms and simplify this to
\begin{multline}
\langle Q \rangle \frac{\partial \alpha_\perp }{\partial Q} + \langle Q \cos^2 \theta \rangle \frac{\partial \Delta \alpha }{\partial Q} = -\frac{1}{4\mu\OV} \left\{  \left(\frac{\partial \alpha}{\partial Q}\right)^2 \sum_{j} \rho^0_{0,j,0,j}  \int_{-\infty}^t \sin[\omega_{1,j,0,j}(t'-t)]E^2 dt'  \right. \\
\left. + \left(\frac{\partial \Delta \alpha}{\partial Q} \right)^2 \sum_{j} \left(\frac{4}{45} \frac{j+1}{(2j-1)(2j+3)} \right)  \rho^0_{0,j,0,j} \int_{-\infty}^t \sin[\omega_{1,j,0,j}(t'-t)]E^2 dt' \right. \\
\left. +  \left(\frac{\partial \Delta \alpha}{\partial Q} \right)^2 \sum_{j} \frac{2}{15} \frac{j(j-1)}{2j-1} \left[ \frac{\rho^0_{0,j-2,0,j-2}}{2j-3}\int_{-\infty}^t \sin[\omega_{1,j,0,j-2}(t'-t)]E^2 dt' \right. \right. \\ \left. \left. -  \frac{\rho^0_{0,j,0,j}}{2j+1} \int_{-\infty}^t \sin[\omega_{0,j,1,j-2}(t'-t)]E^2 dt'\right] \right\}.
\end{multline}
In calculating the vibrational polarization (see Eq.~(\ref{totalpol})) for pump polarization parallel to probe, we neglect the terms containing $\partial \Delta \alpha/\partial Q$, because $\partial \alpha/\partial Q$ and $\partial \Da/\partial Q$ are of the same order, and the terms containing $\partial \Da/\partial Q$ contain additional factors on the order of $1/10$.
This leads to
\begin{equation}
\bP_{vib} \approx  -\left[ \frac{N}{4\mu\OV} \left(\frac{\partial \alpha}{\partial Q}\right)^2 \sum_{j} \rho^0_{0,j,0,j}  \int_{-\infty}^t \sin[\omega_{1,j,0,j}(t'-t)]E^2 dt' \right] \bE.
\end{equation}
If we assume that all of the vibrational frequencies connecting various $j$ states $\omega_{1,j,0,j} \approx \OV$, we can use $\sum_j \rho_{0,j,0,j} = 1$, yielding 
\begin{eqnarray}
\bP_{vib} &\approx& - \left[ \frac{N}{4\mu\OV} \left(\frac{\partial \alpha}{\partial Q} \right)^2 \int_{-\infty}^t \sin[\OV(t'-t)]E^2 dt'\right] \bE \\
&\approx& N \langle Q \rangle_t \left(\frac{\partial \alpha}{\partial Q} \right) \bE,
\label{simplified}
\end{eqnarray}
where $\langle Q \rangle_t$ is given by Eq.~(\ref{Qsimplified}).


\begin{thebibliography}{39}
\expandafter\ifx\csname natexlab\endcsname\relax\def\natexlab#1{#1}\fi
\expandafter\ifx\csname bibnamefont\endcsname\relax
  \def\bibnamefont#1{#1}\fi
\expandafter\ifx\csname bibfnamefont\endcsname\relax
  \def\bibfnamefont#1{#1}\fi
\expandafter\ifx\csname citenamefont\endcsname\relax
  \def\citenamefont#1{#1}\fi
\expandafter\ifx\csname url\endcsname\relax
  \def\url#1{\texttt{#1}}\fi
\expandafter\ifx\csname urlprefix\endcsname\relax\def\urlprefix{URL }\fi
\providecommand{\bibinfo}[2]{#2}
\providecommand{\eprint}[2][]{\url{#2}}

\bibitem[{\citenamefont{Kim et~al.}(2008)\citenamefont{Kim, Taylor, Glownia,
  and Rodriguez}}]{kim_coherent_2008}
\bibinfo{author}{\bibfnamefont{K.~Y.} \bibnamefont{Kim}},
  \bibinfo{author}{\bibfnamefont{A.~J.} \bibnamefont{Taylor}},
  \bibinfo{author}{\bibfnamefont{J.~H.} \bibnamefont{Glownia}},
  \bibnamefont{and}
  \bibinfo{author}{\bibfnamefont{G.}~\bibnamefont{Rodriguez}},
  \bibinfo{journal}{Nature Photonics} \textbf{\bibinfo{volume}{2}},
  \bibinfo{pages}{605} (\bibinfo{year}{2008}),
  \urlprefix\url{http://www.nature.com/nphoton/journal/v2/n10/abs/nphoton.2008.153.html}.

\bibitem[{\citenamefont{Popmintchev et~al.}(2012)\citenamefont{Popmintchev,
  Chen, Popmintchev, Arpin, Brown, Ališauskas, Andriukaitis, Balčiunas,
  Mücke, Pugzlys et~al.}}]{popmintchev_bright_2012}
\bibinfo{author}{\bibfnamefont{T.}~\bibnamefont{Popmintchev}},
  \bibinfo{author}{\bibfnamefont{M.-C.} \bibnamefont{Chen}},
  \bibinfo{author}{\bibfnamefont{D.}~\bibnamefont{Popmintchev}},
  \bibinfo{author}{\bibfnamefont{P.}~\bibnamefont{Arpin}},
  \bibinfo{author}{\bibfnamefont{S.}~\bibnamefont{Brown}},
  \bibinfo{author}{\bibfnamefont{S.}~\bibnamefont{Ališauskas}},
  \bibinfo{author}{\bibfnamefont{G.}~\bibnamefont{Andriukaitis}},
  \bibinfo{author}{\bibfnamefont{T.}~\bibnamefont{Balčiunas}},
  \bibinfo{author}{\bibfnamefont{O.~D.} \bibnamefont{Mücke}},
  \bibinfo{author}{\bibfnamefont{A.}~\bibnamefont{Pugzlys}},
  \bibnamefont{et~al.}, \bibinfo{journal}{Science}
  \textbf{\bibinfo{volume}{336}}, \bibinfo{pages}{1287} (\bibinfo{year}{2012}),
  \urlprefix\url{http://www.sciencemag.org/content/336/6086/1287}.

\bibitem[{\citenamefont{Couairon and
  Mysyrowicz}(2007)}]{couairon_femtosecond_2007}
\bibinfo{author}{\bibfnamefont{A.}~\bibnamefont{Couairon}} \bibnamefont{and}
  \bibinfo{author}{\bibfnamefont{A.}~\bibnamefont{Mysyrowicz}},
  \bibinfo{journal}{Phys. Rep.} \textbf{\bibinfo{volume}{441}},
  \bibinfo{pages}{47} (\bibinfo{year}{2007}).

\bibitem[{\citenamefont{Bergé et~al.}(2007)\citenamefont{Bergé, Skupin,
  Nuter, Kasparian, and Wolf}}]{berge_ultrashort_2007}
\bibinfo{author}{\bibfnamefont{L.}~\bibnamefont{Bergé}},
  \bibinfo{author}{\bibfnamefont{S.}~\bibnamefont{Skupin}},
  \bibinfo{author}{\bibfnamefont{R.}~\bibnamefont{Nuter}},
  \bibinfo{author}{\bibfnamefont{J.}~\bibnamefont{Kasparian}},
  \bibnamefont{and} \bibinfo{author}{\bibfnamefont{J.-P.} \bibnamefont{Wolf}},
  \bibinfo{journal}{Reports on Progress in Physics}
  \textbf{\bibinfo{volume}{70}}, \bibinfo{pages}{1633} (\bibinfo{year}{2007}),
  \urlprefix\url{http://stacks.iop.org/0034-4885/70/i=10/a=R03}.

\bibitem[{\citenamefont{Kolesik and Moloney}(2014)}]{kolesik_modeling_2014}
\bibinfo{author}{\bibfnamefont{M.}~\bibnamefont{Kolesik}} \bibnamefont{and}
  \bibinfo{author}{\bibfnamefont{J.~V.} \bibnamefont{Moloney}},
  \bibinfo{journal}{Reports on Progress in Physics}
  \textbf{\bibinfo{volume}{77}}, \bibinfo{pages}{016401}
  (\bibinfo{year}{2014}),
  \urlprefix\url{http://iopscience.iop.org/0034-4885/77/1/016401}.

\bibitem[{\citenamefont{Peñano et~al.}(2004)\citenamefont{Peñano, Sprangle,
  Hafizi, Ting, Gordon, and Kapetanakos}}]{penano_propagation_2004}
\bibinfo{author}{\bibfnamefont{J.~R.} \bibnamefont{Peñano}},
  \bibinfo{author}{\bibfnamefont{P.}~\bibnamefont{Sprangle}},
  \bibinfo{author}{\bibfnamefont{B.}~\bibnamefont{Hafizi}},
  \bibinfo{author}{\bibfnamefont{A.}~\bibnamefont{Ting}},
  \bibinfo{author}{\bibfnamefont{D.~F.} \bibnamefont{Gordon}},
  \bibnamefont{and} \bibinfo{author}{\bibfnamefont{C.~A.}
  \bibnamefont{Kapetanakos}}, \bibinfo{journal}{Physics of Plasmas}
  \textbf{\bibinfo{volume}{11}}, \bibinfo{pages}{2865} (\bibinfo{year}{2004}),
  \urlprefix\url{http://link.aip.org/link/PHPAEN/v11/i5/p2865/s1}.

\bibitem[{\citenamefont{Varma et~al.}(2008)\citenamefont{Varma, Chen, and
  Milchberg}}]{varma_trapping_2008}
\bibinfo{author}{\bibfnamefont{S.}~\bibnamefont{Varma}},
  \bibinfo{author}{\bibfnamefont{Y.-H.} \bibnamefont{Chen}}, \bibnamefont{and}
  \bibinfo{author}{\bibfnamefont{H.~M.} \bibnamefont{Milchberg}},
  \bibinfo{journal}{Physical Review Letters} \textbf{\bibinfo{volume}{101}},
  \bibinfo{pages}{205001} (\bibinfo{year}{2008}),
  \urlprefix\url{http://link.aps.org/doi/10.1103/PhysRevLett.101.205001}.

\bibitem[{\citenamefont{Chen et~al.}(2010)\citenamefont{Chen, Varma, Antonsen,
  and Milchberg}}]{chen_direct_2010}
\bibinfo{author}{\bibfnamefont{Y.-H.} \bibnamefont{Chen}},
  \bibinfo{author}{\bibfnamefont{S.}~\bibnamefont{Varma}},
  \bibinfo{author}{\bibfnamefont{T.~M.} \bibnamefont{Antonsen}},
  \bibnamefont{and} \bibinfo{author}{\bibfnamefont{H.~M.}
  \bibnamefont{Milchberg}}, \bibinfo{journal}{Physical Review Letters}
  \textbf{\bibinfo{volume}{105}}, \bibinfo{pages}{215005}
  (\bibinfo{year}{2010}),
  \urlprefix\url{http://link.aps.org/doi/10.1103/PhysRevLett.105.215005}.

\bibitem[{\citenamefont{Palastro et~al.}(2012)\citenamefont{Palastro, Antonsen,
  and Milchberg}}]{palastro_compression_2012}
\bibinfo{author}{\bibfnamefont{J.~P.} \bibnamefont{Palastro}},
  \bibinfo{author}{\bibfnamefont{T.~M.} \bibnamefont{Antonsen}},
  \bibnamefont{and} \bibinfo{author}{\bibfnamefont{H.~M.}
  \bibnamefont{Milchberg}}, \bibinfo{journal}{Physical Review A}
  \textbf{\bibinfo{volume}{86}}, \bibinfo{pages}{033834}
  (\bibinfo{year}{2012}),
  \urlprefix\url{http://link.aps.org/doi/10.1103/PhysRevA.86.033834}.

\bibitem[{\citenamefont{Nibbering et~al.}(1997)\citenamefont{Nibbering,
  Grillon, Franco, Prade, and Mysyrowicz}}]{nibbering_determination_1997}
\bibinfo{author}{\bibfnamefont{E.~T.~J.} \bibnamefont{Nibbering}},
  \bibinfo{author}{\bibfnamefont{G.}~\bibnamefont{Grillon}},
  \bibinfo{author}{\bibfnamefont{M.~A.} \bibnamefont{Franco}},
  \bibinfo{author}{\bibfnamefont{B.~S.} \bibnamefont{Prade}}, \bibnamefont{and}
  \bibinfo{author}{\bibfnamefont{A.}~\bibnamefont{Mysyrowicz}},
  \bibinfo{journal}{Journal of the Optical Society of America B}
  \textbf{\bibinfo{volume}{14}}, \bibinfo{pages}{650} (\bibinfo{year}{1997}),
  \urlprefix\url{http://josab.osa.org/abstract.cfm?URI=josab-14-3-650}.

\bibitem[{\citenamefont{Ripoche et~al.}(1997)\citenamefont{Ripoche, Grillon,
  Prade, Franco, Nibbering, Lange, and
  Mysyrowicz}}]{ripoche_determination_1997}
\bibinfo{author}{\bibfnamefont{J.-F.} \bibnamefont{Ripoche}},
  \bibinfo{author}{\bibfnamefont{G.}~\bibnamefont{Grillon}},
  \bibinfo{author}{\bibfnamefont{B.}~\bibnamefont{Prade}},
  \bibinfo{author}{\bibfnamefont{M.}~\bibnamefont{Franco}},
  \bibinfo{author}{\bibfnamefont{E.}~\bibnamefont{Nibbering}},
  \bibinfo{author}{\bibfnamefont{R.}~\bibnamefont{Lange}}, \bibnamefont{and}
  \bibinfo{author}{\bibfnamefont{A.}~\bibnamefont{Mysyrowicz}},
  \bibinfo{journal}{Optics Communications} \textbf{\bibinfo{volume}{135}},
  \bibinfo{pages}{310} (\bibinfo{year}{1997}),
  \urlprefix\url{http://www.sciencedirect.com/science/article/B6TVF-3SPCPK4-89/2/7dffcbff8bf33ce95c1fbcc5d47e787a}.

\bibitem[{\citenamefont{Liu and Chin}(2005)}]{liu_direct_2005}
\bibinfo{author}{\bibfnamefont{W.}~\bibnamefont{Liu}} \bibnamefont{and}
  \bibinfo{author}{\bibfnamefont{S.}~\bibnamefont{Chin}},
  \bibinfo{journal}{Optics Express} \textbf{\bibinfo{volume}{13}},
  \bibinfo{pages}{5750} (\bibinfo{year}{2005}),
  \urlprefix\url{http://www.opticsexpress.org/abstract.cfm?URI=oe-13-15-5750}.

\bibitem[{\citenamefont{Chen et~al.}(2007{\natexlab{a}})\citenamefont{Chen,
  Varma, York, and Milchberg}}]{chen_single-shot_2007}
\bibinfo{author}{\bibfnamefont{Y.-H.} \bibnamefont{Chen}},
  \bibinfo{author}{\bibfnamefont{S.}~\bibnamefont{Varma}},
  \bibinfo{author}{\bibfnamefont{A.}~\bibnamefont{York}}, \bibnamefont{and}
  \bibinfo{author}{\bibfnamefont{H.~M.} \bibnamefont{Milchberg}},
  \bibinfo{journal}{Optics Express} \textbf{\bibinfo{volume}{15}},
  \bibinfo{pages}{11341} (\bibinfo{year}{2007}{\natexlab{a}}),
  \urlprefix\url{http://www.opticsexpress.org/abstract.cfm?URI=oe-15-18-11341}.

\bibitem[{\citenamefont{Odhner et~al.}(2009)\citenamefont{Odhner, Romanov, and
  Levis}}]{odhner_rovibrational_2009}
\bibinfo{author}{\bibfnamefont{J.~H.} \bibnamefont{Odhner}},
  \bibinfo{author}{\bibfnamefont{D.~A.} \bibnamefont{Romanov}},
  \bibnamefont{and} \bibinfo{author}{\bibfnamefont{R.~J.} \bibnamefont{Levis}},
  \bibinfo{journal}{Physical Review Letters} \textbf{\bibinfo{volume}{103}},
  \bibinfo{pages}{075005} (\bibinfo{year}{2009}),
  \urlprefix\url{http://link.aps.org/doi/10.1103/PhysRevLett.103.075005}.

\bibitem[{\citenamefont{Odhner et~al.}(2011)\citenamefont{Odhner, McCole, and
  Levis}}]{odhner_filament-driven_2011}
\bibinfo{author}{\bibfnamefont{J.~H.} \bibnamefont{Odhner}},
  \bibinfo{author}{\bibfnamefont{E.~T.} \bibnamefont{McCole}},
  \bibnamefont{and} \bibinfo{author}{\bibfnamefont{R.~J.} \bibnamefont{Levis}},
  \bibinfo{journal}{The Journal of Physical Chemistry A}
  \textbf{\bibinfo{volume}{115}}, \bibinfo{pages}{13407}
  (\bibinfo{year}{2011}),
  \urlprefix\url{http://dx.doi.org/10.1021/jp207253z}.

\bibitem[{\citenamefont{Kim et~al.}(2002)\citenamefont{Kim, Alexeev, and
  Milchberg}}]{kim_single-shot_2002-1}
\bibinfo{author}{\bibfnamefont{K.~Y.} \bibnamefont{Kim}},
  \bibinfo{author}{\bibfnamefont{I.}~\bibnamefont{Alexeev}}, \bibnamefont{and}
  \bibinfo{author}{\bibfnamefont{H.~M.} \bibnamefont{Milchberg}},
  \bibinfo{journal}{Applied Physics Letters} \textbf{\bibinfo{volume}{81}},
  \bibinfo{pages}{4124} (\bibinfo{year}{2002}).

\bibitem[{\citenamefont{Dogariu et~al.}(1997)\citenamefont{Dogariu, Xia, Hagan,
  Said, Van~Stryland, and Bloembergen}}]{dogariu_purely_1997}
\bibinfo{author}{\bibfnamefont{A.}~\bibnamefont{Dogariu}},
  \bibinfo{author}{\bibfnamefont{T.}~\bibnamefont{Xia}},
  \bibinfo{author}{\bibfnamefont{D.~J.} \bibnamefont{Hagan}},
  \bibinfo{author}{\bibfnamefont{A.~A.} \bibnamefont{Said}},
  \bibinfo{author}{\bibfnamefont{E.~W.} \bibnamefont{Van~Stryland}},
  \bibnamefont{and}
  \bibinfo{author}{\bibfnamefont{N.}~\bibnamefont{Bloembergen}},
  \bibinfo{journal}{Journal of the Optical Society of America B}
  \textbf{\bibinfo{volume}{14}}, \bibinfo{pages}{796} (\bibinfo{year}{1997}),
  \urlprefix\url{http://josab.osa.org/abstract.cfm?URI=josab-14-4-796}.

\bibitem[{\citenamefont{Smolorz and Wise}(2000)}]{smolorz_femtosecond_2000}
\bibinfo{author}{\bibfnamefont{S.}~\bibnamefont{Smolorz}} \bibnamefont{and}
  \bibinfo{author}{\bibfnamefont{F.}~\bibnamefont{Wise}},
  \bibinfo{journal}{Journal of the Optical Society of America B}
  \textbf{\bibinfo{volume}{17}}, \bibinfo{pages}{1636} (\bibinfo{year}{2000}),
  \urlprefix\url{http://josab.osa.org/abstract.cfm?URI=josab-17-9-1636}.

\bibitem[{\citenamefont{Wahlstrand et~al.}(2013)\citenamefont{Wahlstrand,
  Odhner, McCole, Cheng, Palastro, Levis, and
  Milchberg}}]{wahlstrand_effect_2013}
\bibinfo{author}{\bibfnamefont{J.~K.} \bibnamefont{Wahlstrand}},
  \bibinfo{author}{\bibfnamefont{J.~H.} \bibnamefont{Odhner}},
  \bibinfo{author}{\bibfnamefont{E.~T.} \bibnamefont{McCole}},
  \bibinfo{author}{\bibfnamefont{Y.-H.} \bibnamefont{Cheng}},
  \bibinfo{author}{\bibfnamefont{J.~P.} \bibnamefont{Palastro}},
  \bibinfo{author}{\bibfnamefont{R.~J.} \bibnamefont{Levis}}, \bibnamefont{and}
  \bibinfo{author}{\bibfnamefont{H.~M.} \bibnamefont{Milchberg}},
  \bibinfo{journal}{Physical Review A} \textbf{\bibinfo{volume}{87}},
  \bibinfo{pages}{053801} (\bibinfo{year}{2013}),
  \urlprefix\url{http://link.aps.org/doi/10.1103/PhysRevA.87.053801}.

\bibitem[{\citenamefont{Laban et~al.}(2010)\citenamefont{Laban, Wallace,
  Glover, Sang, and Kielpinski}}]{laban_self-focusing_2010}
\bibinfo{author}{\bibfnamefont{D.~E.} \bibnamefont{Laban}},
  \bibinfo{author}{\bibfnamefont{W.~C.} \bibnamefont{Wallace}},
  \bibinfo{author}{\bibfnamefont{R.~D.} \bibnamefont{Glover}},
  \bibinfo{author}{\bibfnamefont{R.~T.} \bibnamefont{Sang}}, \bibnamefont{and}
  \bibinfo{author}{\bibfnamefont{D.}~\bibnamefont{Kielpinski}},
  \bibinfo{journal}{Optics Letters} \textbf{\bibinfo{volume}{35}},
  \bibinfo{pages}{1653} (\bibinfo{year}{2010}),
  \urlprefix\url{http://ol.osa.org/abstract.cfm?URI=ol-35-10-1653}.

\bibitem[{\citenamefont{Wahlstrand
  et~al.}(2011{\natexlab{a}})\citenamefont{Wahlstrand, Cheng, Chen, and
  Milchberg}}]{wahlstrand_optical_2011-1}
\bibinfo{author}{\bibfnamefont{J.~K.} \bibnamefont{Wahlstrand}},
  \bibinfo{author}{\bibfnamefont{Y.-H.} \bibnamefont{Cheng}},
  \bibinfo{author}{\bibfnamefont{Y.-H.} \bibnamefont{Chen}}, \bibnamefont{and}
  \bibinfo{author}{\bibfnamefont{H.~M.} \bibnamefont{Milchberg}},
  \bibinfo{journal}{Physical Review Letters} \textbf{\bibinfo{volume}{107}},
  \bibinfo{pages}{103901} (\bibinfo{year}{2011}{\natexlab{a}}),
  \urlprefix\url{http://link.aps.org/doi/10.1103/PhysRevLett.107.103901}.

\bibitem[{\citenamefont{Wahlstrand
  et~al.}(2012{\natexlab{a}})\citenamefont{Wahlstrand, Cheng, and
  Milchberg}}]{wahlstrand_absolute_2012}
\bibinfo{author}{\bibfnamefont{J.~K.} \bibnamefont{Wahlstrand}},
  \bibinfo{author}{\bibfnamefont{Y.-H.} \bibnamefont{Cheng}}, \bibnamefont{and}
  \bibinfo{author}{\bibfnamefont{H.~M.} \bibnamefont{Milchberg}},
  \bibinfo{journal}{Physical Review A} \textbf{\bibinfo{volume}{85}},
  \bibinfo{pages}{043820} (\bibinfo{year}{2012}{\natexlab{a}}),
  \urlprefix\url{http://link.aps.org/doi/10.1103/PhysRevA.85.043820}.

\bibitem[{\citenamefont{Wahlstrand
  et~al.}(2012{\natexlab{b}})\citenamefont{Wahlstrand, Cheng, and
  Milchberg}}]{wahlstrand_high_2012}
\bibinfo{author}{\bibfnamefont{J.~K.} \bibnamefont{Wahlstrand}},
  \bibinfo{author}{\bibfnamefont{Y.-H.} \bibnamefont{Cheng}}, \bibnamefont{and}
  \bibinfo{author}{\bibfnamefont{H.~M.} \bibnamefont{Milchberg}},
  \bibinfo{journal}{Physical Review Letters} \textbf{\bibinfo{volume}{109}},
  \bibinfo{pages}{113904} (\bibinfo{year}{2012}{\natexlab{b}}),
  \urlprefix\url{http://link.aps.org/doi/10.1103/PhysRevLett.109.113904}.

\bibitem[{\citenamefont{Loriot et~al.}(2009)\citenamefont{Loriot, Hertz,
  Faucher, and Lavorel}}]{loriot_measurement_2009}
\bibinfo{author}{\bibfnamefont{V.}~\bibnamefont{Loriot}},
  \bibinfo{author}{\bibfnamefont{E.}~\bibnamefont{Hertz}},
  \bibinfo{author}{\bibfnamefont{O.}~\bibnamefont{Faucher}}, \bibnamefont{and}
  \bibinfo{author}{\bibfnamefont{B.}~\bibnamefont{Lavorel}},
  \bibinfo{journal}{Optics Express} \textbf{\bibinfo{volume}{17}},
  \bibinfo{pages}{13429} (\bibinfo{year}{2009}),
  \urlprefix\url{http://www.opticsexpress.org/abstract.cfm?URI=oe-17-16-13429}.

\bibitem[{\citenamefont{Reichert et~al.}(2015)\citenamefont{Reichert, Zhao,
  Reed, Ensley, Hagan, and Van~Stryland}}]{reichert_beam_2015}
\bibinfo{author}{\bibfnamefont{M.}~\bibnamefont{Reichert}},
  \bibinfo{author}{\bibfnamefont{P.}~\bibnamefont{Zhao}},
  \bibinfo{author}{\bibfnamefont{J.~M.} \bibnamefont{Reed}},
  \bibinfo{author}{\bibfnamefont{T.~R.} \bibnamefont{Ensley}},
  \bibinfo{author}{\bibfnamefont{D.~J.} \bibnamefont{Hagan}}, \bibnamefont{and}
  \bibinfo{author}{\bibfnamefont{E.~W.} \bibnamefont{Van~Stryland}},
  \bibinfo{journal}{Optics Express} \textbf{\bibinfo{volume}{23}},
  \bibinfo{pages}{22224} (\bibinfo{year}{2015}), 
  \urlprefix\url{https://www.osapublishing.org/abstract.cfm?URI=oe-23-17-22224}.

\bibitem[{\citenamefont{Chen et~al.}(2007{\natexlab{b}})\citenamefont{Chen,
  Varma, Alexeev, and Milchberg}}]{chen_measurement_2007}
\bibinfo{author}{\bibfnamefont{Y.-H.} \bibnamefont{Chen}},
  \bibinfo{author}{\bibfnamefont{S.}~\bibnamefont{Varma}},
  \bibinfo{author}{\bibfnamefont{I.}~\bibnamefont{Alexeev}}, \bibnamefont{and}
  \bibinfo{author}{\bibfnamefont{H.}~\bibnamefont{Milchberg}},
  \bibinfo{journal}{Optics Express} \textbf{\bibinfo{volume}{15}},
  \bibinfo{pages}{7458} (\bibinfo{year}{2007}{\natexlab{b}}),
  \urlprefix\url{http://www.opticsexpress.org/abstract.cfm?URI=oe-15-12-7458}.

\bibitem[{\citenamefont{Irikura}(2007)}]{irikura_experimental_2007}
\bibinfo{author}{\bibfnamefont{K.~K.} \bibnamefont{Irikura}},
  \bibinfo{journal}{Journal of Physical and Chemical Reference Data}
  \textbf{\bibinfo{volume}{36}}, \bibinfo{pages}{389} (\bibinfo{year}{2007}),
  \urlprefix\url{http://scitation.aip.org/content/aip/journal/jpcrd/36/2/10.1063/1.2436891}.

\bibitem[{\citenamefont{Huber and Herzberg}(1979)}]{huber_molecular_1979}
\bibinfo{author}{\bibfnamefont{K.~P.} \bibnamefont{Huber}} \bibnamefont{and}
  \bibinfo{author}{\bibfnamefont{G.}~\bibnamefont{Herzberg}},
  \emph{\bibinfo{title}{Molecular {Spectra} and {Molecular} {Structure}: {IV}.
  {Constants} of {Diatomic} {Molecules}}} (\bibinfo{publisher}{van Nostrand
  Reinhold}, \bibinfo{address}{New York}, \bibinfo{year}{1979}).

\bibitem[{\citenamefont{Zahedpour et~al.}(2014)\citenamefont{Zahedpour,
  Wahlstrand, and Milchberg}}]{zahedpour_quantum_2014}
\bibinfo{author}{\bibfnamefont{S.}~\bibnamefont{Zahedpour}},
  \bibinfo{author}{\bibfnamefont{J.}~\bibnamefont{Wahlstrand}},
  \bibnamefont{and}
  \bibinfo{author}{\bibfnamefont{H.}~\bibnamefont{Milchberg}},
  \bibinfo{journal}{Physical Review Letters} \textbf{\bibinfo{volume}{112}},
  \bibinfo{pages}{143601} (\bibinfo{year}{2014}),
  \urlprefix\url{http://link.aps.org/doi/10.1103/PhysRevLett.112.143601}.

\bibitem[{\citenamefont{Bridge and
  Buckingham}(1966)}]{bridge_polarization_1966}
\bibinfo{author}{\bibfnamefont{N.~J.} \bibnamefont{Bridge}} \bibnamefont{and}
  \bibinfo{author}{\bibfnamefont{A.~D.} \bibnamefont{Buckingham}},
  \bibinfo{journal}{Proceedings of the Royal Society of London. Series A.
  Mathematical and Physical Sciences} \textbf{\bibinfo{volume}{295}},
  \bibinfo{pages}{334} (\bibinfo{year}{1966}),
  \urlprefix\url{http://rspa.royalsocietypublishing.org/content/295/1442/334}.

\bibitem[{\citenamefont{Bishop and Pipin}(1993)}]{bishop_vibrational_1993}
\bibinfo{author}{\bibfnamefont{D.~M.} \bibnamefont{Bishop}} \bibnamefont{and}
  \bibinfo{author}{\bibfnamefont{J.}~\bibnamefont{Pipin}},
  \bibinfo{journal}{The Journal of Chemical Physics}
  \textbf{\bibinfo{volume}{98}}, \bibinfo{pages}{522} (\bibinfo{year}{1993}),
  \urlprefix\url{http://scitation.aip.org/content/aip/journal/jcp/98/1/10.1063/1.465057}.

\bibitem[{\citenamefont{Demtroder}(2005)}]{demtroder_molecular_2005}
\bibinfo{author}{\bibfnamefont{W.}~\bibnamefont{Demtroder}},
  \emph{\bibinfo{title}{Molecular {Physics}}} (\bibinfo{publisher}{Wiley-VCH},
  \bibinfo{year}{2005}).

\bibitem[{\citenamefont{Wahlstrand and
  Milchberg}(2011)}]{wahlstrand_effect_2011}
\bibinfo{author}{\bibfnamefont{J.~K.} \bibnamefont{Wahlstrand}}
  \bibnamefont{and} \bibinfo{author}{\bibfnamefont{H.~M.}
  \bibnamefont{Milchberg}}, \bibinfo{journal}{Optics Letters}
  \textbf{\bibinfo{volume}{36}}, \bibinfo{pages}{3822} (\bibinfo{year}{2011}),
  \urlprefix\url{http://ol.osa.org/abstract.cfm?URI=ol-36-19-3822}.

\bibitem[{\citenamefont{Lundeen et~al.}(1983)\citenamefont{Lundeen, Hou, and
  Nibler}}]{lundeen_nonresonant_1983}
\bibinfo{author}{\bibfnamefont{T.}~\bibnamefont{Lundeen}},
  \bibinfo{author}{\bibfnamefont{S.-Y.} \bibnamefont{Hou}}, \bibnamefont{and}
  \bibinfo{author}{\bibfnamefont{J.~W.} \bibnamefont{Nibler}},
  \bibinfo{journal}{The Journal of Chemical Physics}
  \textbf{\bibinfo{volume}{79}}, \bibinfo{pages}{6301} (\bibinfo{year}{1983}),
  \urlprefix\url{http://scitation.aip.org/content/aip/journal/jcp/79/12/10.1063/1.445736}.

\bibitem[{\citenamefont{Shelton and Rice}(1994)}]{shelton_measurements_1994}
\bibinfo{author}{\bibfnamefont{D.~P.} \bibnamefont{Shelton}} \bibnamefont{and}
  \bibinfo{author}{\bibfnamefont{J.~E.} \bibnamefont{Rice}},
  \bibinfo{journal}{Chem. Rev.} \textbf{\bibinfo{volume}{94}},
  \bibinfo{pages}{3} (\bibinfo{year}{1994}),
  \urlprefix\url{http://dx.doi.org/10.1021/cr00025a001}.

\bibitem[{\citenamefont{Shelton}(1990)}]{shelton_nonlinear-optical_1990}
\bibinfo{author}{\bibfnamefont{D.~P.} \bibnamefont{Shelton}},
  \bibinfo{journal}{Physical Review A} \textbf{\bibinfo{volume}{42}},
  \bibinfo{pages}{2578} (\bibinfo{year}{1990}),
  \urlprefix\url{http://link.aps.org/doi/10.1103/PhysRevA.42.2578}.

\bibitem[{\citenamefont{Rosasco and Hurst}(1986)}]{rosasco_dispersion_1986}
\bibinfo{author}{\bibfnamefont{G.~J.} \bibnamefont{Rosasco}} \bibnamefont{and}
  \bibinfo{author}{\bibfnamefont{W.~S.} \bibnamefont{Hurst}},
  \bibinfo{journal}{Journal of the Optical Society of America B}
  \textbf{\bibinfo{volume}{3}}, \bibinfo{pages}{1251} (\bibinfo{year}{1986}),
  \urlprefix\url{https://www.osapublishing.org/josab/abstract.cfm?uri=josab-3-10-1251}.

\bibitem[{\citenamefont{Hahn and Lee}(1995)}]{hahn_measurement_1995}
\bibinfo{author}{\bibfnamefont{J.~W.} \bibnamefont{Hahn}} \bibnamefont{and}
  \bibinfo{author}{\bibfnamefont{E.~S.} \bibnamefont{Lee}},
  \bibinfo{journal}{Journal of the Optical Society of America B}
  \textbf{\bibinfo{volume}{12}}, \bibinfo{pages}{1021} (\bibinfo{year}{1995}),
  \urlprefix\url{https://www.osapublishing.org/josab/abstract.cfm?uri=josab-12-6-1021}.

\end{thebibliography}

\end{document}